\documentclass[11pt]{article}\usepackage[dvips]{graphicx}
\usepackage{latexsym}
\usepackage{amssymb}
\usepackage{bm}

\newcommand{\C}{\mathbb{C}}

\usepackage{amsmath, amsthm, amssymb}

\def\bea{\begin{eqnarray}}
\def\eea{\end{eqnarray}}

\def\be{\begin{equation}}
\def\ee{\end{equation}}

\def\TD{{\mathcal{D}}} 
\def\cs{\sigma}
\def\cv{\nu}
\def\cg{\mathbf{g}}

\begin{document}
\title{\vskip -70pt
\vskip 60pt
{\bf Summary of session A4: Complex and conformal methods in classical and quantum gravity}
\vskip 10pt}
\author{Maciej Dunajski\thanks{\tt m.dunajski@damtp.cam.ac.uk}
\\[8pt]
{\sl Department of Applied Mathematics and Theoretical Physics} \\
{\sl University of Cambridge} \\
{\sl Wilberforce Road, Cambridge CB3 0WA, UK} \\[8pt]}
\date{} 
\maketitle
\begin{abstract} 
This paper summarises oral contributions to the
parallel session {\it Complex and conformal methods in classical and quantum gravity} which took place during the  20th International Conference on General Relativity and Gravitation held in Warsaw in July 2013.
\end{abstract}
\section{Introduction}
The methods of complex analysis and conformal geometry in general relativity go back to the research on gravitational radiation and global, asymptotic structures of space-times done in 1960s. 
Twistor theory of Roger Penrose combines both complex and conformal methods in an attempt to unify gravity and quantum theory in a way which makes the light-rays more fundamental than space-time points.
Over the years this approach has influenced research in pure mathematics, 
ranging from differential geometry of hyper-Kahler manifolds, to an overview of integrable systems arising from the self-duality equations
on Yang-Mills connection or conformal structure.

The topics discussed in the talks split into three categories:
{\em Twistor theory} covers applications of the twistor methods to cosmology,
perturbative scattering amplitudes, as well as quantum gravity and non-commutative twistors.
{\em Conformal methods} contains presentations concerned either with applications of conformal Einstein field equations, or Einstein--Weyl geometry. Finally {\em Complex methods, algebraically special metrics
and gravitational radiation} is devoted to complex solutions to Einstein equations ($H$ and $HH$ spaces), as well as real algebraically special solutions resulting form the existence of shear-free geodesics 
congruences.

\section{The contributions}
\subsection{Twistor Theory}
In his talk  {\em Cosmological Twistors} {\bf Roger Penrose} has described a  connection between
the conformal cyclic cosmology (CCC), and hyper-surface twistors.
According to CCC
the universe consists of an unending sequence of
aeons, each beginning with a conformally
smooth big bang and continuing to an exponential
expansion whose conformal infinity joins smoothly
(as a conformal 4-manifold) to the conformally
expanded big-bang of the succeeding aeon \cite{penrose_CCC}. Each
crossover 3-surface ${\mathcal{X}}_r$ is a spacelike conformal 3-manifold (owing to positive  $\Lambda$), and is associated
with a symplectic 6-manifold ${\mathcal{N}}_r$, each of whose
points represents a scaled null-geodesic segment
intersecting ${\mathcal{X}}_r$  (the scaling being provided by a
null covector ${\bf p}$, parallel-propagated and pointing
along it, this being a conformally invariant
characterization). The real 6-space  ${\mathcal{N}}_r$ is closely
associated with the hypersurface twistor space of
${\mathcal{X}}_r$.
These null geodesics which do not extend into
black-hole singularities can be continued to the
succeeding ${\mathcal{X}}_{r+1}$ , thereby providing a symplectic
map of  ${\mathcal{N}}_r$ into  ${\mathcal{N}}_{r+1}$. At first sight, this appears to
provide an argument that the symplectic volumes
of the ${\mathcal{N}}_r$  must continually decrease as $r$ increases,
in contradiction with the ideas of CCC (this
seeming to be a particularly manifest issue when  ${\mathcal{X}}_r$
 is compact). However, this apparent difficulty is
resolved by the loss of null geodesics in black holes
being accommodated by the (non-compact) scaling
change for null geodesics that pass near to the
black holes without entering them.
\vskip5pt

In his second contribution to the Session
{\bf Roger Penrose} presented his ideas on  {\em Non-commutative
twistor space and curved space-time}.
The standard Non-Linear Graviton construction, in
\v{C}ech form, involves the `gluing' different open
patches of twistor space together in order to
construct a `curved' twistor space \cite{penrose_NG}. The required
complex smoothness of the gluing, may be
expressed in a matching together of the sheaf of
holomorphic functions over each patch. This gives
rise to the construction of anti-self-dual complex
space-times.
A novel approach to generalizing this procedure,
giving hope for the possibility of removing the
anti-self-dual restriction, arose from a recent brief
conversation between  Roger Penrose and Michael Atiyah. The idea is to
generalize the earlier construction in a way that
would `match', in a certain sense, the sheaves of
holomorphic functions, {\it without} there being a
corresponding matching of the underlying twistor
spaces. To achieve what seems to be needed, one
must match not the sheaves of holomorphic
functions but, instead, the (completed) twistor
Heisenberg algebras, so as to provide a `non-
commutative twistor geometry'. This includes, as
special cases, both the twistor and dual twistor
patchings of the Non-Linear Graviton construction.
In recent work, following the delivery of this talk,
the ideas of geometric quantization have emerged
as probably providing the next step towards what is
needed.	
\vskip5pt
Another approach to non-commutative twistor theory was given by
{\bf Jerzy Lukierski} in his presentation
{\em Noncommutative Space-Time From Quantized Twistors}.
Following original Penrose's twistorial framework one can consider the
relativistic phase space coordinates $(x_{\mu },p_{\mu })$ as composite,
described by functions of the primary pair $t_{A}^{i},\bar{t}_{A}^{i}$ $%
(A=1...4,i=1,2)$ of twistor components. The canonical twistor 
Poisson brackets (PB)
imply the deformation of
relativistic phase space PB algebra and this leads, after quantization, to
noncommutative composite space-time coordinates. The resulting deformed
relativistic PB can be closed as nonlinear (nonpolynomial) PB algebra if one
supplements the Pauli-Lubanski four-vector $w_{\mu }$ (such that $p_{\mu }w^{\mu }=0$) by what leads to new spin-extended deformed Heisenberg algebra. In order to
quantize the PB algebra with generators $(x_{\mu },p_{\mu },w_{\mu })$ \ one
should use the star-product quantization method of Kontsevich, which
will provide the associative quantum spin-extended deformed Heisenberg
algebra. Such spin-extended phase space can be linked with the relativistic
symplectic structures which characterize the Poincare orbits for spinning
massive particles  in agreement with the earlier works from 1970s of
Souriau,  Kunzle and Tod.
\vskip5pt
{\bf Simone Speziale} gave a talk
{\it Loop quantum gravity, twisted geometries and twistors}.
Loop quantum gravity is a background-independent approach to the quantization of general relativity. It is often useful to consider a truncation thereof, defined on the lattice dual to a graph. This truncation captures a finite number of degrees of freedom, which have been shown to describe a certain generalization of Regge geometries, called twisted geometries \cite{Freidel:2010aq}. These discrete geometries can be described in terms of a collection of twistors associated to the graph (see \cite{Speziale:2012nu} and references therein). The relation between twistors and loop quantum gravity hinges on the fact that the holonomy-flux algebra, the basic algebra of observables in loop quantum gravity, can be described in terms of twistors.

The symplectic space $T^*SL(2,\C)$, with its canonical Poisson algebra, can be derived from an Hamiltonian reduction of the phase space of a pair of twistors, constrained to have the same complex helicity. 
The twistor incidence relation coincides with the time-like vector used in the 3 + 1 splitting of the gravitational action.
Once holonomies and fluxes are parametrized in terms of twistors, the $SU(2)$ gauge invariance around each node of a spin network translates into the
condition that the
space-like planes, identified by the restricted incidence relation, close
to form a flat bounded, convex
polyhedron, with areas of the faces given by the 
twistor helicity.
This classical description of the phase space of loop quantum gravity on a fixed graph can be quantized in a canonical way. 
\vskip5pt
{\bf Lionel Mason's} talk
{\it Gravity in twistor space}
surveyed the recent progress in understanding gravity in twistor space, particularly in regard to the computation of scattering amplitudes.  Gravity can be obtained from twistor theory in two distinct ways.  One is to
consider a string theory in twistor space, and the other is to try to construct an Einstein gravity action in twistor space.  The twistor-string approach started with Witten's twistor-string theory in 2003.  This theory described
conformal supergravity, i.e., a supersymmetric extension of the theory whose  action is the square of the Weyl tensor.  There is a closely related twistor action that plays the role of an effective action for the string theory.
Much more recently, a twistor-string formula for the gravity MHV amplitude due to Hodges, was extended to the whole tree-level S-matrix by Cachazo and Skinner, and proved by Cachazo, Mason and  Skinner \cite{cms}.  Subsequently, Skinner found
an underlying twistor-string model that gave rise to the formula.  More recent work of Adamo and Mason \cite{am}, based on an argument of Maldacena concerning the use of conformal gravity to compute Einstein amplitudes with cosmological
constant,  use twistor-strings and twistor-actions for conformal gravity to obtain the Einstein MHV amplitude with a cosmological constant.  This leads to a conjecture for the Einstein twistor action with cosmological constant.

\vskip5pt
{\bf Sergiu Vacaru} gave a talk
{\it Twistors and Almost K\"{a}hler Models  of Einstein and Finsler Spaces}.
He described a  generalization of the spinor and twistor geometry for Finsler-Cartan spaces modeled on tangent Lorentz bundles, or on (pseudo) Riemannian manifolds \cite{v1}. Nonholonomic (Finsler) twistors are defined as solutions of generalized twistor equations determined by spin connections and frames adapted to a nonlinear connection structure. The constructions for local twistors can be globalized using nonholonomic deformations with auxiliar metric compatible connections completely determined by the metric structure and/or the Finsler fundamental function. The theory of nonholonomic spinors and Dirac operators \cite{v2} 
can be applied in order to encode gravitational field equations into almost K\"{a}hler twistor distributions. Such an approach can be applied in the Einstein gravity theory formulated in Finsler like variables for a nonholonomic 2+2 splitting. This allows to generate using twistor methods new classes of generic off--diagonal exact solutions in (modified) Einstein and Finsler gravity theories.

\vskip5pt
{\bf George Sparling's} presentation {\it Quantum Twistor Theory and the Big Bang}
described his joint work with a research student Jonathan Holland 
\cite{sparling}. The authors claim that
homogeneous cosmological models are unstable under small perturbations. 
They construct a  stable (in their sense)  family of perturbed metrics involving a  scalar field, which is driving the conformal expansion of 
the universe and obeying 
the  sine-Gordon equation.

\subsection{Conformal methods}
In the talk {\em On the conformal structure of the extremal
Reissner-Nordstr\"om spacetime} {\bf Juan Antonio Valiente Kroon}
presented his work with Christian L\"ubbe.
The conformal Einstein field equations 
have been used with remarkable success to understand the existence and
stability of {asymptotically simple spacetimes}. It is natural to ask whether it is
possible to adapt these ideas to analyse the stability of black hole
spacetimes. One of the underlying strategies is to obtain a detailed understanding of the geometric
structure of the background solution under consideration in order to construct 
an evolution problem which is as simple as possible. In this respect, the conformal structure of the
extremal Reissner-Nordstr\"om seems particularly amenable to a
detailed analysis. 
It has been shown \cite{LueVal13b}  that the
domain of outer communication of the extremal Reissner-Nordstr\"om
spacetime can be covered by a non-intersecting congruence of curves
with special conformal properties. These curves are of special interest as they provide a
simple expression for a conformal factor which, in turn, could be used
to obtain a conformal representation of the spacetime. Moreover,
these curves can be used as the cornerstone of a gauge for the
conformal evolution equations. 
In \cite{LueVal13c}  the discrete conformal
isometry of the extremal Reissner-Nordstr\"om spacetime has been used 
to obtain a
representation of timelike infinity for which the conformal field
equations and their associated initial data are regular. This construction
may have
implications for the propagation of test fields
and non-linear perturbations of the gravitational field close to the horizon.
\vskip5pt
{\bf  Christian L\"ubbe~}
explained his results  in the talk
{\it The Einstein field equations for the conformal scalar field in higher
dimensions}.
Given a conformal manifold $(M, [\cg])$
of dimension greater than two  
there exists a 1-1 correspondence between the conformal scale $\cs$ of the Einstein metric $\tilde{g} \in [\cg]$ and a class of constant sections $Y$ of the tractor bundle over $(M, [\cg])$
\cite{GoverAlmostEinstein}. By fixing a conformal scale $\cv$ the conformal Einstein field equations (CEFE) \cite{FriCEFE2002} are recovered.
For the conformal scale $\cs$ of a physical metric with matter the above approach gives rise to $\TD Y = {\mathcal{T}}$ for some non-vanishing tractor $ {\mathcal{T}}$. Generally ${\mathcal{T}}$ would be constructed by hand from the energy-momentum tensor. L\"ubbe  explained that for the conformal scalar field it is actually possible to construct $ {\mathcal{T}}$ in a canonical way. The construction works for any choice of cosmological constant $\Lambda$ and permissible source term for the conformal scalar field. Moreover, one obtains a conformally invariant, symmetric, trace-free and divergence free-tensor $T_{ij}$, which can be interpreted as the energy-momentum tensor of the conformal scalar field with source term. The decomposition of $T_{ij}$ is identical in all dimensions and the coefficients of different components depend on the dimension. Fixing a conformal scale $\cv $ one can once more read off the CEFE  from the tractor equations. 
It remains to be seen whether other matter models can be treated in this way.
\vskip5pt
{\bf Jan Gutowski} gave a talk
{\it Supersymmetry and Einstein--Weyl Structures}
based on his work with Dunajski, Sabra, Tod \cite{ew2} and
Klemm, Sabra and Sloane \cite{ew3}.
Three dimensional Einstein-Weyl structures arise in four and five dimensional supergravity theories in several contexts. 
A $n$-dimensional Einstein-Weyl manifold $M$ admits a metric $g$, 
a torsionless connection $D$, and a 1-form ${\cal{B}}$, satisfying
\begin{equation*}
Dg= 2 {\cal{B}} \otimes g, \qquad W_{(ab)} = {1 \over n}W_c{}^c g_{ab} \ ,
\end{equation*}
where  $W_{(ab)}$ denotes the symmetrized Ricci tensor of the connection $D$. 
Einstein-Weyl geometry appears in a particularly natural fashion in four-dimensional gauged Euclidean supergravity. The bosonic sector of
this theory is Einstein-Maxwell theory with a cosmological constant.
Supersymmetric solutions also admit a Killing spinor, satisfying a first order Killing spinor equation.
Using spinorial geometry techniques, $Spin(4)$ gauge transformations can be used to write the Killing spinor in a particularly simple canonical form \cite{ew2}. The Killing spinor equation can then be explicitly solved, and the 4-dimensional metric is
\begin{equation*}
ds^2 = {1 \over 2} \bigg({2 \sqrt{2} \ell \over \sigma^2-1}d \sigma
-{\ell \over \sqrt{2}} (\sigma^{-1}+\sigma) {\cal{B}}+(\sigma^{-1}-\sigma) \xi \bigg)^2+{\sigma^2 \over (1-\sigma^2)^2}ds^2_{M} \ ,
\end{equation*}
where $\ell$ is constant, $ds^2_M$ denotes the metric on a 3-dimensional Einstein-Weyl manifold $M$, and ${\cal{B}}$, $\xi$ are 1-forms on $M$ satisfying
\begin{equation*}
d {\cal{B}}=-\ell^{-1} \star_M {\cal{B}}, \qquad  d \xi + {\cal{B}} \wedge \xi =-\ell^{-1} \star_M \xi \ .
\end{equation*}

Einstein-Weyl structures also appear in the near-horizon geometries of
extremal supersymmetric five-dimensional black holes \cite{ew3}. In this case, the Killing spinor equations are analysed by writing the geometry in Gaussian null co-ordinates. The five-dimensional metric is then given by
\begin{equation*}
ds^2=2 du \bigg(dr+r {\cal{B}}-{1 \over 2}r^2 \Phi^2 du\bigg)+ds^2_{M},
\qquad d \Phi + \Phi {\cal{B}} + \star_M d {\cal{B}}=0 \ ,
\end{equation*}
where again $ds^2_M$ is the metric on a 3-dimensional Einstein-Weyl manifold $M$, ${\cal{B}}$ and $\Phi$ are a 1-form and a scalar field on $M$ respectively. This structure is present not only for near-horizon black hole geometries in the minimal five-dimensional theory, but also
in significantly more complicated higher derivative five-dimensional supergravities.
\vskip5pt
In his talk
{\it Einstein-Weyl geometry and webs}
{\bf Wojciech Kry\'nski} described his work with Maciej Dunajski
 \cite{DK} on  solutions to Einstein-Weyl equations on three 
dimensional manifolds. The solutions are constructed from special families of foliations, called Veronese webs, which were introduced 
by Gelfand and Zakharevich as reductions of bi-Hamiltonian structures. It appears that the Einstein-Weyl structures of the hyper-CR type are locally given in terms of solutions to the dispersionless Hirota equation.  Kry\'nski has also
discussed connections between Einstein-Weyl geometry and third order 
ODEs \cite{DK2}  and some generalisations in other dimensions.

\subsection{Complex methods, algebraically special metrics and gravitational radiation.}
{\bf Daniel Finley}  gave a talk 
{\it CR Structures Allow a New Second-Order ode to Determine Some Twisting Type N Einstein Spaces}
based on the work with his research student Zhang \cite{finley}.
Algebraically special, twisting Einstein spaces admit a 3-parameter
congruence of shearfree null geodesics,
which maps to a CR structure determined uniquely by the metric and the
congruence.
The equivalence problem for CR structures, solved by Cartan, depends on
invariants determined  by
one complex function.  One can aim to use this  structure to  
search for
solutions of Petrov Type $N$ with non-zero twist, which are
incredibly rare.
 Insisting on the existence of a 
a Killing vector made the problem solvable, although it required
the cosmological constant not to vanish.  The
characteristics for the classical symmetries were determined, and the {\it group-invariant} solutions were
found leaving only
a single ODE to determine the complete solution, with one
constant of integration. If this constant vanishes then 
the ODE is
the Abel equation
which could not so far be solved.
A diffent ODE arises if the constant of integration is non-zero.
Finley has presented a detailed analysis of this equation, and recovered some known solutions.
The general solution appears to be given by transcendental function which is however well behaved on the real axis.
\vskip5pt
In the presentation
{\it Complex and spinorial approaches to higher-dimensional
general relativity} {\bf Arman Taghavi-Chabert} sumarised results of his recent works \cite{arman1, arman2}.
He introduced higher-dimensional generalisations of the notions of {shearfree congruences of null geodesics} and the {Petrov-Penrose classification of the Weyl tensor}, both of which were intrumental to the discovery of exact solutions of Einstein's equations, such as the Kerr metric and the Robinson-Trautman metric. The generalisations impinge on the fact that on four-dimensional {complexified} spacetime, a shearfree congruence of null geodesics can be interpreted as a complex holomorphic foliation tangent to a totally null complex $2$-plane distribution.

Thus, a natural arena is a complex Riemannian manifold $(\mathcal{M},g)$ of dimension $2m$ equipped with a totally null $m$-plane distribution $\mathcal{N}$, or equivalently, a \emph{pure} spinor field (up to scale) $\xi$. Pointwise, the stabiliser $P$, say, of either $\mathcal{N}$ or the line $\langle \xi \rangle$ of pure spinors, is a parabolic Lie subgroup of the complex special orthogonal group $\mathrm{SO}(2m,\mathbb{C})$. The representation theory of $P$ 
can be used
to classify tensor fields on $(\mathcal{M},g)$ irreducibly. This leads in particular to the pointwise algebraic classifications of 
\begin{itemize}
\item the tensor $\nabla \langle \xi \rangle$, where $\nabla$ is the Levi-Civita connection on $\mathcal{M}$ -- algebraic conditions on $\nabla \langle \xi \rangle$ can then be interpreted as geometric conditions on $\mathcal{N}$, eg. integrability...;
\item the Weyl tensor, thereby generalising Penrose's notion of principal null spinors of the Weyl tensor, and, to some extent, the Petrov-Penrose classification of the (self-dual) Weyl tensor.
\end{itemize}
Within this framework, one can define an algebraically special condition on the Weyl tensor, dependent on $\xi$, which leads to a partial analogue of the Goldberg-Sachs theorem in higher dimensions, i.e. to the integrability of the totally null $m$-plane distribution associated to $\xi$.

\vskip5pt
In his talk {\it Partner symmetries, group foliation and lift of non-invariant
solutions of heavenly equations from 3 to 4 dimensions},
{\bf Mikhail Sheftel} presented results of his joint work with Andriey Malykh.
Non-invariant solutions of complex Monge-Amp\`ere equation ($CMA$), with no
symmetry reduction in the number of independent variables, generate
anti-self-dual Ricci-flat metrics of Euclidean signature without Killing
vectors  \cite{pleban} which is one of characteristic features of the $K3$ 
instanton.

Sheftel demonstrated how to construct  noninvariant
solutions of the $CMA$  and provide a lift from rotationally invariant solutions
of $CMA$ satisfying Boyer-Finley equation to non-invariant ones.
The symmetry condition for $CMA$ has a two-dimensional divergence form and
determines locally a potential which is also a symmetry.
The original symmetry and its potential, called \textit{partner symmetries},
are related by a non-local recursion relation.
The symmetry group parameters can then be introduced as additional variables. The
recursion relation  provides
differential constraints compatible with the $CMA$. To simplify the
equations,  symmetry reductions involving group parameters are used instead of 
 reductions in original variables, thus avoiding invariant solutions.
To solve the  equations,  the  \textit{group foliation}
method \cite{ns,foliat} is employed. This leads to  
new non-invariant solutions of the $CMA$
equation
together with anti-self-dual Ricci-flat metric without Killing vectors. The
metric is free of singularities but the curvature is not concentrated in a
bounded domain, so it is not an instanton metric.
\vskip5pt
{\bf Adam Chudecki} talked about his work on 
{\it Null Killing and null homothetic Killing vectors 
and their connection with geometry of null strings.}
The existence of null isometric and null homothetic Killing vectors have 
an impact on
the structure of the complex Einstein spaces. These vectors define totally
null and geodesic, self-dual and anti-self-dual 2-surfaces, the null strings
(twistor surfaces) \cite{dunajski_west}. The Goldberg-Sachs theorem then implies that both
self-dual and anti-self-dual parts of the Weyl tensor of such spaces are
algebraically degenerate. Consequently, the spaces considered are
hyperheavenly spaces ($HH$-spaces) or, if one of the parts of the Weyl tensor
vanishes, heavenly spaces ($H$-spaces).
Existence of null homothetic Killing vectors implies 
the vanishing of the cosmological
constant. The only spaces, which admit such symmetries, are hyperheavenly
spaces of the types [III]x[N] or heavenly spaces of the types [III,N]x[-].
The null strings which are defined by the null homothetic Killing vectors
have interesting property: if the self-dual congruence of the null strings is
expanding then anti-self-dual congruence of the null strings is necesarilly
nonexpanding or vice-versa. Moreover, type [III] always corresponds to the
nonexpanding congruence of the null strings, while type [N] corresponds to
the expanding congruence. Heavenly metrics admitting null homothetic Killing
vectors can all be found explicitly. The field equations for the
hyperheavenly space of the type [III]x[N] can been reduced to the single
nonlinear partial differential equation. Complex Einstein spaces equipped
with the null homothetic Killing vector do not admit any Lorentzian slices.

Metrics of the complex Einstein spaces admitting null isometric Killing
vector have been presented explicitly, except spaces of the type [II]x[II].
In the case of the type [II]x[II] with $\Lambda = 0$, Einstein field equations
have been reduced to the Euler-Poisson-Darboux equation. Lorentzian slices
of the types [II]x[II], [D]x[D] and [N]x[N] have been found. Examples of the
Einstein spaces with the neutral signature have been presented, one of the
most interesting is the most general, globally Osserman but not Walker
space, equipped with the null isometric Killing vector \cite{chudecki}.

\vskip5pt
In the talk 
{\it Newton's 2nd Law,  Radiation  Reaction, and Type $II$   Einstein-Maxwell 
Fields} {\bf Ted Newman} described his work on algebraicaly special solutions. 
  In analogy with referring to the source of a Schwarzschild, Reissner-Nordstrom or Kerr-Newman metric as a Schwarzschild,
Reissner-Nordstrom or Kerr-Newman particle, one can refer to the sources of the type $II$ metrics as type $II$ particles. The field equations are then interpreted as the equations governing the dynamics (motion and multipole behavior)
of the type $II$ particles. The metric itself is not given here.
   If one starts with the full Einstein-Maxwell equations and imposes the type $II$ conditions many of the resulting equations (the radial equations) can be easily integrated, leaving four (complex) reduced type $II$ Einstein-Maxwell
equations - (non-linear and rather intractable looking) - for four complex field variables, which are functions of time and two angles, plus a reality condition. Two of the four dependent variables are spin-coefficient versions of
the Maxwell field, one of them is a spin-coefficient Weyl tensor component and the last is a (geometric) direction field that specifies the direction (at null infinity) of the shear-free null geodesic congruence defined by the
type $II$ condition. The structure of the four field equations is as follows: the first two equations and the third determine respectively, the two Maxwell fields and the Weyl tensor component (that contains the Bondi
energy-momentum four-vector) in terms of the direction field. The last equation, basically a conservation law, is the dynamics for the direction field.
   Working within the algebraically special type $II$ Einstein-Maxwell field equations, one considers perturbations from the Reissner-Nordstrom metric: 
  linear perturbations (for the first three equations) with
spherical harmonic expansions up to and including the $l=2$ terms.  The fourth of the field equations, (the $l=0,1$ terms form the Bondi energy-momentum conservation law) is intrinsically quadratic and is essentially empty if
linearized. With no inconsistency, the results obtained from the three linearized equations are inserted into the fourth. The main results are in the $l=0$ and $1$ harmonics of the last equation.  The $l=0$ is the energy conservation law containing both gravitational quadrupole radiation and electric and magnetic
dipole and quadrupole radiation of a charged particle in complete agreement with known classical results. The $l=1$ terms yield Newtons $2^{nd}$  law of motion, with the force coming from both radiation reaction terms and
momentum recoil terms.   The $l=2$ terms determine the evolution of the quadrupole moments. These results do
not involve any model building and do not contain any mass renormalization procedure.

\vskip5pt

In the talk {\it Spin and Center of Mass in Asymptotically Flat Spacetimes}
{\bf Carlos Kozameh} presented his results obtained with Gonzalo Quiroga.
The work  defined the notion of center of mass and spin for asymptotically flat spacetimes, i.e., spacetimes where there is a precise notion of an isolated gravitational system \cite{KQ}. The main tools used in the construction are the Geroch-Winicour linkages \cite{GW} together with a canonical foliation constructed from worldlines in the spacetime.
Equations of motion for these variables linking their time evolution to the emitted gravitational radiation were found. 
In astrophysics one often assumes conservation laws for isolated systems. However, the analysis suggests show that for highly energetic processes were a fair percentage of energy is emitted as gravitational radiation, this is far from being true. The equations give an explicit algebraic relationship between the gravitational radiation and the center of mass acceleration and time rate of the spin. The solutions to the above equations yield their dynamical evolution.
In particular, these equations should be useful in the description of the late stage of closed binary coalescence, after a black hole has formed.


\begin{thebibliography}{jafsdl}
\frenchspacing


\bibitem{penrose_CCC} Penrose, R. (2010)  
{\em Cycles of Time: An Extraordinary New View of the Universe}, 
Bodley Head.


\bibitem{penrose_NG} Penrose, R. (1976) Nonlinear 
gravitons and curved twistor theory, Gen. Rel. Grav.  {\bf 7},  31--52.


\bibitem{Freidel:2010aq} Freidel, L.  and Speziale, S.
(2010)
Twisted geometries: A geometric parametrisation of SU(2) phase space
Phys.\ Rev.\ D {\bf 82}, 084040.

\bibitem{Speziale:2012nu}
Speziale, S.  and Wieland, W. W. (2012)
The twistorial structure of loop-gravity transition amplitudes,
Phys.\ Rev.\ D {\bf 86},  124023.



\bibitem{am} Adamo, T. and Mason, L. J. (2013)
Conformal and Einstein gravity from twistor actions 
{\tt arXiv:1307.5043}.

\bibitem{cms} Cachazo, F., Mason, L. J. and Skinner, D. (2012)
Gravity in Twistor Space and its Grassmannian Formulation.
{\tt arXiv:1207.4712}




\bibitem{v1}  Vacaru, S. (2012) Finsler Spinors and Twistors in Einstein Gravity and Modifications,  {\tt arXiv: 1206.4012}

\bibitem{v2} Vacaru, S. (2009) Spectral functionals, nonholonomic Dirac operators, and noncommutative Ricci flows, J. Math. Phys.  \textbf{50}  073503.




\bibitem{sparling} Holland, J. and Sparling, G. (2013) 
The cosmology of a fundamental scalar {\tt arXiv:1307.3922}.


\bibitem{LueVal13b}
L\"ubbe, C.  \& {Valiente Kroon}, J. A. (2013)
A class of conformal curves in the Reissner-Nordstr{\"o}m
  spacetime,
\newblock to appear in Ann. Henri Poincar\'e. {\tt arXiv1301.5458}

\bibitem{LueVal13c}
L\"ubbe, C.  \& {Valiente Kroon}, J. A. (2013)
On the conformal structure of the extremal
    Reissner-Nordstr\"om spacetime.
    {\tt arXiv1308.1325}

\bibitem{GoverAlmostEinstein}
Gover, A. R. (2004)
{ Almost conformally Einstein manifolds and obstructions}, in \newblock Proceedings of the 9th International Conference on Differential Geometry and its Applications,  {\tt math.DG/0412393}.

\bibitem{FriCEFE2002}
Friedrich, H. (2002)
{Conformal Einstein Evolution}, in 
'The Conformal Structure of Space-Time' Lecture Notes in Physics Volume {\bf 604},  p 1-50 {\tt gr-qc/0209018}.


\bibitem{ew2}
  Dunajski, M.,  Gutowski, J. B, Sabra, W. A,  and Tod, K. P. (2011)
  Cosmological Einstein-Maxwell Instantons and Euclidean Supersymmetry: Beyond Self-Duality,
  J. High Energy Phys. {\bf 03} 131;
  {\tt arXiv:1012.1326}.
  
\bibitem{ew3}
  Gutowski, J. B., Klemm, D.,  Sabra, W. A. and Sloane, P. (2012)
  Small Horizons,
  J. High Energy Phys. {\bf 01} (2012) 146; {\tt arXiv:1109.1566 }.


\bibitem{DK} Dunajski, M. and W. Kry\'nski, W. (2013) 
{Einstein--Weyl geometry, dispersionless Hirota equation and Veronese webs}, {\tt arXiv:1301.0621}.

\bibitem{DK2}
Dunajski M. and Kry\'nski, W. (2013) Point invariants of third-order ODEs and hyper-CR Einstein-Weyl structures {\tt arXiv:1310.5704}.

\bibitem{finley}
Zhang, X. and Finley, D., (2012) Lower-order ODEs to
determine new twisting
type $N$ Einstein
spaces via CR geometry, Class. Quant. Grav. {\bf 29}, 065010.


\bibitem{arman1} Taghavi-Chabert, A.(2012)
{Pure spinors, intrinsic torsion and curvature in even dimensions}. 
{arXiv:1212.3595}

\bibitem{arman2} Taghavi-Chabert, A.(2013)
Pure spinors, intrinsic torsion and curvature in odd dimensions. 
\texttt{arXiv:1304.1076}.




\bibitem{pleban} 
Pleba\'nski, J. (1975). Some solutions of complex Einstein equations. J. Math Phys. {\bf 16}, 2395–2402.


\bibitem{ns}
Nutku Y. and Sheftel M. B., (2001)
Differential invariants and group foliation for the complex Monge-Amp\`ere
equation,
\textit{J. Phys. A:  Math.  Gen.} {\bf 34} 137--156.
\bibitem{foliat}
Martina, L., Sheftel, M. B. and  Winternitz P., (2001)
Group foliation and non-invariant solutions of the heavenly equation,
\textit{J. Phys. A: Math. Gen.} \textbf{34}, 9243--9263.

\bibitem{dunajski_west} 
Dunajski, M. and West, S. (2007), Anti-Self-Dual Conformal Structures with
Null Killing
Vectors from Projective Structures, Commun. Math. Phys. {\bf 272}, 85-118


\bibitem{chudecki} Chudecki, A. (2013) 
Null Killing vectors and geometry of null strings in
Einstein spaces, {\tt arXiv:1306.6216}. 


\bibitem{KQ} Kozameh, C. N. and Quiroga, G. D. (2012)
Spin and Center of Mass in Axially Symmetric Einstein-Maxwell Spacetimes.,
Class. Quantum Grav. {\bf 29}, 235006

\bibitem{GW} Geroch, R. and Winicour, J. (1981) 
Linkages in General Relativity,
J. Math. Phys., {\bf 22}, 803-12


\end{thebibliography}
\end{document}